\documentclass[twocolumn]{aastex631}
\usepackage{amsmath}
\usepackage{amssymb}

\submitjournal{ApJS}

\shorttitle{Energy-conserving Curvilinear PIC}
\shortauthors{J.\ Croonen et al.}

\DeclareMathOperator{\sech}{sech}

\newcommand\bb[1]{\mbox{\boldmath{$#1$}}}

\newcommand\grad{\bb{\nabla}}

\begin{document}

\title{Exactly energy-conserving electromagnetic Particle-in-Cell method in curvilinear coordinates}

\author[0000-0002-3587-116X]{J. Croonen}
\affiliation{Centre for mathematical Plasma Astrophysics, Department of Mathematics, KU Leuven, Celestijnenlaan 200B, B-3001 Leuven, Belgium}

\author[0000-0001-5079-7941]{L. Pezzini}
\affiliation{Centre for mathematical Plasma Astrophysics, Department of Mathematics, KU Leuven, Celestijnenlaan 200B, B-3001 Leuven, Belgium}
\affiliation{Royal Observatory of Belgium, Solar-Terrestrial Centre of Excellence, Ringlaan 3, 1180 Uccle, Belgium}

\author[0000-0002-7526-8154]{F. Bacchini}
\affiliation{Centre for mathematical Plasma Astrophysics, Department of Mathematics, KU Leuven, Celestijnenlaan 200B, B-3001 Leuven, Belgium}
\affiliation{Royal Belgian Institute for Space Aeronomy, Solar-Terrestrial Centre of Excellence, Ringlaan 3, 1180 Uccle, Belgium}

\author[0000-0002-3123-4024]{G.Lapenta}
\affiliation{Centre for mathematical Plasma Astrophysics, Department of Mathematics, KU Leuven, Celestijnenlaan 200B, B-3001 Leuven, Belgium}
 
\begin{abstract}
In this paper, we introduce and discuss an exactly energy-conserving Particle-in-Cell method for arbitrary curvilinear coordinates. The flexibility provided by curvilinear coordinates enables the study of plasmas in complex-shaped domains by aligning the grid to the given geometry, or by focusing grid resolution on regions of interest without overresolving the surrounding, potentially uninteresting domain. 
We have achieved this through the introduction of the metric tensor, the Jacobian matrix, and contravariant operators combined with an energy-conserving fully implicit solver. We demonstrate the method's capabilities using a Python implementation to study several one- and two-dimensional test cases: the electrostatic two-stream instability, the electromagnetic Weibel instability, and the geomagnetic environment modeling (GEM) reconnection challenge. The test results confirm the capability of our new method to reproduce theoretical expectations (e.g.\ instability growth rates) and the corresponding results obtained with a Cartesian uniform grid when using curvilinear grids. Simultaneously, we show that the method conserves energy to machine precision in all cases.

\end{abstract}


\section{Introduction} \label{sec:intro}
Advanced plasma simulation methods have become a key tool for studying a wide variety of systems and phenomena throughout the fields of plasma physics and astrophysics. The difficulty of recreating the relevant plasma conditions in a laboratory environment make simulations irreplaceable for obtaining otherwise unachievable key insight. Among plasma simulations techniques, Particle-in-Cell (PIC) methods stand out as particularly useful when studying plasmas where kinetic effects are important. The state of the art of PIC methods is constantly evolving with new capabilities and improvements being developed. Recently, the astro- and plasma communities have steered their interest toward more advanced PIC methods, especially involving flexible (i.e.\ adaptive) grids and high-stability algorithms, to allow for long-term simulations of systems where Cartesian (uniform) grids might be ill-suited (for example Tokamak simulations using magnetic coordinates as shown in e.g.\ \citealt{Jolliet_2007}, or accretion disks around astrophysical compact objects using spherical Kerr-Schild coordinates as shown in e.g.\ \citealt{Crinquand_2022}). A nonuniform, adaptive grid could be particularly beneficial in several cases, e.g.\ i) when the physical shape of the domain of interest is nontrivial, such that a significant part of an encompassing Cartesian box would be occupied by empty space or otherwise uninteresting regions where plasma behavior is trivial or quasi-static; and ii) when a small region of interest is embedded in a much larger domain, and the latter might be significantly overresolved (in terms of spatial and temporal resolution) due to physical requirements only imposed by the small region of interests. In both cases, employing a standard uniform grid can substantially increase the computational cost of the simulation, in the worst case making certain studies entirely unfeasible. The introduction of a nonuniform grid is one way to relax these limitations, by adapting the grid to the physical setup in question and thus assigning the computational resources more efficiently (see e.g.\ \citealt{Fichtl_2012, delzanno_2013, chacon_2016, Stanier_2022}). 

Several methods have been developed to use curvilinear grids, but most are implemented with one specific grid in mind, and thus lack flexibility (see e.g.\ \citealt{Ringle_2013,Gonzalez_2019}). Others are fully general, i.e.\ they work with an arbitrary grid, but come with other trade-offs such as: i) using a reduced set of equations, rather than a full electromagnetic implementation, thus limiting their applicability to the constraints of these approximations (e.g.\ an electrostatic model in \citealt{Fichtl_2012, delzanno_2013}, or the Vlasov-Darwin approximation in \citealt{chacon_2016}); or ii) introducing large deviations from energy conservation in long simulations, due to discretization errors that are intrinsic of the PIC implementation with explicit methods. Energy conservation is a universal physical property, of particular interest in setups where energy is converted from one form to another. For example, when modeling instabilities or reconnection events, preserving energy is extremely important to avoid numerical heating or cooling which could significantly influence the plasma dynamics and potentially invalidate the results (e.g.\ \citealt{Markidis_2011}).

In this work, we provide a method combining general curvilinear grids, through the introduction of coordinate transformations, with a fully electromagnetic-PIC implementation using an exactly energy-conserving, Jacobian-free Newton-Krylov solver. The method was designed with the architecture of the \textsc{ECsim} code (\citealt{Lapenta_Ecsim}) in mind, to allow for easy implementation in a production-ready infrastructure. To the best of our knowledge, this may represent the first production-scale code which combines the aforementioned properties (curvilinear grids and exact energy conservation). 

This paper is organized as follows: the mathematical foundation of the method is described in detail in Section~\ref{sec:method}; in Section~\ref{sec:tests}, a series of one- and two-dimensional tests are discussed to assess the validity of the new method; the main results, conclusions, potential applications, and next steps in this line of research are discussed in the final Section~\ref{sec:discussion}.

\section{The Curvilinear PIC Method}
\label{sec:method}

The method described in this paper combines a fully implicit PIC implementation (\citealt{chen2011energy,markidis2011energy}) using finite differences and a Jacobian-free Newton-Krylov solver, with a curvilinear grid through the introduction of the metric tensor, Jacobian matrix, and covariant differentiation operators. As we will show in Section~\ref{sec:tests}, the method was successfully implemented and tested in Python.

\subsection{Governing equations}
The PIC method is a first-principles approach for fully kinetic plasma simulations. It describes the plasma as a coupled system between freely moving particles existing in a position-velocity phase space, and a discrete representation of electric and magnetic fields on a computational grid.

To evolve the fields in time we use Maxwell's equations, here in CGS units:
\begin{equation}
\label{eq:divE=rho}
    \grad\cdot \boldsymbol{E} = 4\pi\rho, 
\end{equation}
\begin{equation}
    \grad\cdot \boldsymbol{B} = 0, \label{eq:divB=0}
\end{equation}
\begin{equation}
    \frac{1}{c}\frac{\partial\boldsymbol{B}}{\partial t} = -\grad\times\boldsymbol{E},\label{eq:max1}
\end{equation}
\begin{equation}
    \frac{1}{c}\frac{\partial\boldsymbol{E}}{\partial t} = -\frac{4\pi}{c} \boldsymbol{I} + \grad\times\boldsymbol{B},\label{eq:max2}
\end{equation}
where $\boldsymbol{E}$ and $\boldsymbol{B}$ are the electric and magnetic field respectively, $\rho$ is the charge density, $c$ is the speed of light, and $\boldsymbol{I}$ is the current density, which is computed from the particle velocities (see Section~\ref{sec:discr} below). Note that the PIC method only requires Faraday's law~\eqref{eq:max1} and Amp\`ere's law~\eqref{eq:max2} to form a closed system of equations. It can easily be shown that Gauss's law for magnetism \eqref{eq:divB=0} will always be satisfied if the spatial discretization on the computational grid mimics the continuous vector identity $\grad\cdot\grad\times \bb{V} = 0$ for a generic vector $\bb{V}$ \citep{Lapenta_Ecsim}. Gauss's law for $\bb{E}$ \eqref{eq:divE=rho}, however, is not automatically satisfied in fully implicit PIC. While it is possible to construct a method that also satisfies this equation (e.g.\ \citealt{chen2011energy,charge_conserving_pic_2019}), we are currently only concerned with presenting a first implementation of our method, which conserves energy exactly in curvilinear coordinates, and that can than be further improved upon in future work. Moreover, in our tests we did not detect numerical artifacts linked to charge conservation, implying that numerical errors are limited.

Computational particles in the PIC method are updated using the standard equations of motion,
\begin{equation}
\label{eq:newt1}
\frac{\mathrm{d}\boldsymbol{x}}{\mathrm{d}t} = \boldsymbol{v}, 
\end{equation}
\begin{equation}
\label{eq:newt2}
\frac{\mathrm{d}\boldsymbol{v}}{\mathrm{d}t} = \frac{q}{m} \left(\boldsymbol{E} + \frac{1}{c}\boldsymbol{v} \times \boldsymbol{B}\right),
\end{equation}
where $\boldsymbol{x}$ and $\boldsymbol{v}$ are the particle position and velocity, $q$ the particle charge, and $m$ the particle mass.

In the PIC paradigm, the sources for Maxwell's equations ($\boldsymbol{I}$ and $\rho$) are collected from the particles onto the computational grid via interpolation; electromagnetic fields are interpolated at the particle positions to obtain the Lorentz force needed to update the particles. Because of this, particle and field equations are in general nonlinearly coupled. In the method described in this paper a fully implicit nonlinear iterative solver is used to solve this set of equations, which are discretized as described in Section~\ref{sec:discr}.

\subsection{Curvilinear PIC: Coordinate transformation}

\begin{figure}
    \centering
    \includegraphics[width=1\columnwidth, trim={0 0mm 0mm 0}, clip]{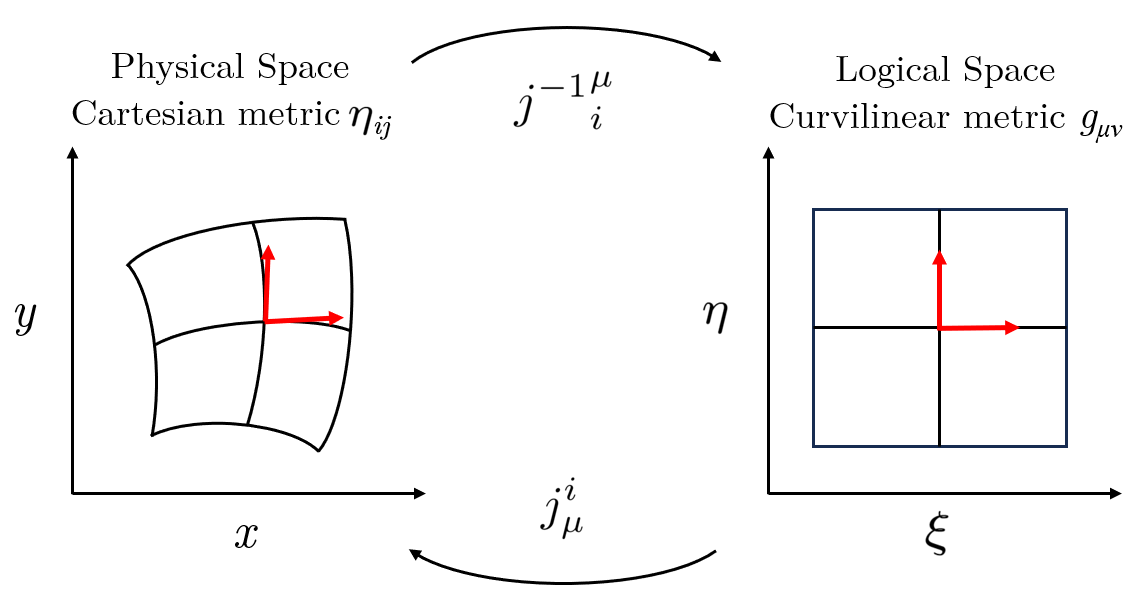}
    \caption{ Diagram of the coordinate transformation between physical and logical space. The Jacobian and its inverse are used to transform contravariant vector components. In physical space the Euclidean metric is used, while a nontrivial metric is used in logical space.}
    \label{fig:diagram}
\end{figure}

To work with a curved grid we consider an arbitrary mapping
\begin{equation}\label{eq:f}
    \boldsymbol{\xi} = f(\boldsymbol{x}).
\end{equation}
The space described by the new coordinates $\boldsymbol{\xi}$ will be called the ``logical'' space and the space described by the original set of coordinates $\boldsymbol{x}$ will be called the ``physical'' space. The mapping $f$ is chosen such that the curvilinear grid in physical space becomes a regular orthogonal grid in logical space. This is depicted in figure \ref{fig:diagram}. Next, we introduce the Jacobian and its inverse,
\begin{equation}
j_{\mu}^{i} = \frac{\partial x_i}{\partial \xi_{\mu}},
\end{equation}
\begin{equation}
{j^{-1}}_i^\mu = \frac{\partial \xi_{\mu}}{\partial x_i},
\end{equation}
and the associated metric defined as
\begin{equation}\label{eq:metric}
    g_{\mu\nu} = j^{i}_{\mu}j^{i}_{\nu}.
\end{equation}
Here, Greek indices are used to denote $\boldsymbol{\xi}$ coordinates in the logical space and Latin indices to denote $\boldsymbol{x}$ coordinates in physical space. In this way, the difficulty of working with a nonuniform curvilinear grid has been circumvented by transforming to the logical grid, at the cost of a non-trivial metric tensor and Jacobian matrix. Calculating the metric and Jacobian only has to be done once at initialization for each grid point and can subsequently be stored in memory. Note that, in our implementation, the grid is static and therefore the Jacobian and metric are time-invariant.

Due to the introduction of a nontrivial metric the vector operators have to be changed from their usual Cartesian implementation to a covariant implementation. The only differential operator appearing in our PIC equations is the curl operator, e.g.\ for a generic (covariant) vector field $V_\kappa$,
\begin{equation}
    (\grad \times \bb{V})^{\mu} \rightarrow \frac{1}{J}\epsilon^{\mu\nu\kappa}\partial_\nu V_{\kappa},
\end{equation}
where $J=|j|$ is the Jacobian determinant, and $\epsilon^{\mu\nu\kappa}$ is the Levi-Civita symbol. Furthermore, the metric tensor can be used to raise and lower indices, e.g.\ $V_{\kappa} = g_{\kappa\lambda}V^{\lambda}$. 

For the particle equations of motion we keep the Cartesian description from eqs.~\eqref{eq:newt1}--\eqref{eq:newt2}. We avoid solving these equations in logical space, since this would require the Jacobian and metric tensor to be known at each particle position at every timestep. Given the computational complexity of deriving these values from the initial mapping function, this was not considered a practical operation to be carried out at every iteration for every particle. This operation is in principle feasible if all mapping quantities are analytically known, but this imposes additional constraints on the mapping function (i.e.\ invertibility and differentiability) which we want to avoid for generality. Calculating the equations of motion in Cartesian coordinates instead introduces the need to convert particle positions and vector-field components between physical and logical space. This strategy is still less computationally demanding than evolving particles in the logical space. Particle positions are transformed from Cartesian physical space to general logical space simply using the original mapping function $f$ in eq.~\eqref{eq:f}. Contravariant vector components can be transformed using the Jacobian and inverse Jacobian matrices as
\begin{equation}
V^{i} = j_{\mu}^{i}V^{\mu}, \label{eq:gen2cart}    
\end{equation}
\begin{equation}
V^{\mu} = {j^{-1}}_i^\mu V^{i}.\label{eq:cart2gen}   
\end{equation}
    
This is necessary in the interpolation of the current density and the fields, as we show in the following Section. For further details or interest in the topic we recommend the book \cite{liseikin1999grid}.

\subsection{Discretization}
\label{sec:discr}
The notation in this paper uses superscript parentheses to denote discrete time steps, and subscript parentheses to denote discrete spatial coordinates. All quantities are known at integer time steps; when the method requires half-integer times (see below), the relevant quantities will be linearly interpolated between subsequent time levels. Field quantities are defined at discrete locations on the computational grid, e.g.\ in two spatial dimensions, $V_{(\xi,\eta)}^{(n)}$ denotes a quantity $V$ at the grid location $(\xi,\eta)$ in the logical space and discrete time level $n$. This allows us to avoid confusion with indices of vector components. When the exact spatial indices are not relevant they are replaced by subscript $g$ to indicate a quantity known at a generic grid point, or subscript $p$ for a quantity known at a particle's position.

To discretize the governing equations, we use a finite-difference scheme. Magnetic-field components are placed on grid nodes, with integer spatial indices, while electric field components are placed on the grid centers, with half-integer spatial indices\footnote{This choice conforms to the ``colocated'' grid discretization that has been used for the \textsc{iPic3D} and \textsc{ECSIM} family of codes, see e.g.\ \cite{markidis2010,Lapenta_Ecsim}.}. Maxwell's equations~\eqref{eq:max1}--\eqref{eq:max2} require spatial derivatives of the magnetic field to be colocated with the electric field, and conversely the spatial derivatives of the electric field to be colocated with the magnetic field. This requires the quantities to be averaged along the direction perpendicular to that of the derivative before taking finite differences. This allows for a centered finite-difference spatial derivative which is second-order accurate; in two dimensions, the discrete derivatives for a generic quantity $V$ along the $\xi$ or $\eta$ directions are
\begin{equation}
    D_{\xi} V = \frac{(V_{(\xi+1,\eta)}+V_{(\xi+1,\eta+1)}) - (V_{(\xi,\eta)} + V_{(\xi,\eta+1)})}{2\Delta\xi},
\end{equation}
\begin{equation}
    D_{\eta} V = \frac{(V_{(\xi,\eta+1)}+V_{(\xi+1,\eta+1)}) - (V_{(\xi,\eta)} + V_{(\xi+1,\eta)})}{2\Delta\eta},
\end{equation}
where the spatial indices can be integers or half-integers depending on whether the specific quantity exists on nodes or centers respectively. Time derivatives are discretized using a second-order accurate central-difference scheme. Using these discrete derivatives, we can now write the discrete Maxwell's equations \eqref{eq:max1}--\eqref{eq:max2}: 
\begin{equation}
    \frac{B^\mu{}_g^{(n+1)}-B^\mu{}_g^{(n)}}{\Delta t} = - \frac{c}{J_g}\epsilon^{\mu\nu\kappa} (D_{\nu}E_{\kappa})_g^{(n+1/2)},
    \label{eq:max1disc}
\end{equation}
\begin{equation}
\begin{split}
    \frac{E^\mu{}_g^{(n+1)}-E^\mu{}_g^{(n)}}{\Delta t} = &-4\pi I^{\mu}{}_g^{(n+1/2)} \\ &+ \frac{c}{J_g}\epsilon^{\mu\nu\kappa} (D_{\nu}B_{\kappa})_g^{(n+1/2)}.
    \label{eq:max2disc}
\end{split}
\end{equation}
The current density on each grid point must be gathered from the particles as
\begin{equation}\label{eq:I}
    I^{\mu}{}_g =  \frac{({j^{-1}}_i^\mu)_g}{J_g\Delta\xi\Delta\eta} \sum_p w_{pg}q_{p}v^{i}{}_{p},
\end{equation}
where $w_{pg}$ is the interpolation function linking particle $p$ with the grid point $g$. $w_{pg}$ is a function of the particle position in logical coordinates; this can be found using the mapping function $\boldsymbol{\xi} = f(\boldsymbol{x})$. The inverse-Jacobian factor in the equation above transforms the current from physical to logical components as described by eq.~\eqref{eq:cart2gen}. 

Since particles are evolved in the physical space the discretization is straightforward, 
\begin{equation}
    \frac{x^{i}{}_p^{(n+1)} - x^{i}{}_p^{(n)}}{\Delta t} = v^{i}{}_p^{(n+1/2)}, \label{eq:newt1disc}
\end{equation}
\begin{equation}
    \frac{v^{i}{}_p^{(n+1)} - v^{i}{}_p^{(n)}}{\Delta t} = \frac{q_p}{m_p} (E^{i} + \epsilon^{ijk}v_{j}B_{k})^{(n+1/2)}_p. \label{eq:newt2disc}
\end{equation}
Note that the fields in these equations are located at the particle positions, which requires an interpolation: for a generic field $V$,
\begin{equation}\label{eq:grid_to_part}
    V^{i}{}_p = \sum_g w_{gp} (j_{\mu}^{i}V^{\mu})_g ,
\end{equation}
where $w_{gp}$ is the interpolation function from grid points to particle positions. By construction we have chosen $w_{gp} = w_{pg}$, which is a requirement for energy conservation (see Appendix~\ref{apx:nrg}). As described by eq.~\eqref{eq:gen2cart}, we added the Jacobian $j_{\mu}^{i}$ to convert $V^{\mu}$ from curvilinear to Cartesian coordinates, as required by the Cartesian equations of motion.

To solve for equations~\eqref{eq:max1disc}--\eqref{eq:max2disc} and \eqref{eq:newt1disc}--\eqref{eq:newt2disc}, a nonlinear Newton-Krylov iterative solver is used. Convergence of this solver is typically case-based, and is not guaranteed when spatiotemporal resolution is insufficient to capture relevant physical phenomena. As a rule of thumb for the tests in the next Sections, the temporal and spatial resolutions were constrained to satisfy $c\Delta t / \Delta x < 1$, with a typical value for this ratio usually set to $\sim 0.25$. This was sufficient in all test cases to achieve absolute and relative errors below $10^{-14}$, typically within $\sim$10 iterations. Note that $\Delta x$ may vary throughout our nonuniform grids, in which case the smallest value of $\Delta x$ in the grid must be considered for this constraint.

\subsection{Discrete energy}
To verify the exactly energy-conserving nature of this method, a proper definition for the relevant energies must be established.
The electric- and magnetic-field energies are respectively defined as
\begin{equation}
    U_\mathrm{E} = \sum_g \frac{\Delta\xi\Delta\eta}{8\pi}(J E_\mu E^{\mu})_g,\label{eq:Ue}
\end{equation}
\begin{equation}
    U_\mathrm{B} = \sum_g \frac{\Delta\xi\Delta\eta}{8\pi}(J B_{\mu}B^{\mu})_g,\label{eq:Ub}
\end{equation}
which combined give the total electromagnetic-field energy $U_\mathrm{F} = U_\mathrm{E} + U_\mathrm{B}$. The particle kinetic energy is defined as
\begin{equation}\label{eq:Uk}
    U_\mathrm{K} = \sum_p\frac{m (v_{i}v^i)_p}{2}.
\end{equation}
The sum of the kinetic and electromagnetic field energy is the total energy in the system: $U = U_\mathrm{K} + U_\mathrm{F}$. In Appendix~\ref{apx:nrg} we show analytically that, using these definitions, the total energy will be conserved exactly (i.e.\ to machine precision). 

\section{Test cases}\label{sec:tests}
To validate the methods developed in this paper, we considered three different test cases: the one-dimensional, electrostatic two-stream instability; the one-dimensional, electromagnetic Weibel (or filamentation) instability \citep{weibel, bret2010}; and the two-dimensional Geospace Environmental Modeling (GEM) challenge, i.e.\ a paradigmatic magnetic-reconnection setup \citep{GEM}. 

\subsection{Two-stream Instability}
\label{sbsec:2stream}

The two-stream instability represents a classical one-dimensional test for PIC codes. The initial setup consists of two counterstreaming electron beams with uniform number density $n_0/2$. We initialize the particles in each beam according to a drifting Maxwellian with thermal speed $v_{\mathrm{th}}/c = 0.001$. The drift-velocity component is  along the only spatial direction of the system $x$, and is added to both beams with opposite sign, i.e.\ $v_{\mathrm{d}}/c = \pm 0.2$. The analytic solution of the dispersion relation predicts that the system is unstable if $k v_\mathrm{d} /\omega_{\mathrm{p}}<1$ \citep{Goldston-1995}, with $k$ the wave number and $\omega_{\mathrm{p}}$ the plasma frequency. This results in an unstable, purely electrostatic evolution where charge bunching will generate a self-reinforcing electric-field perturbation. During the linear instability phase, the electric-field energy (eq.~\eqref{eq:Ue}) in the fastest-growing unstable mode will grow as

\begin{equation}\label{eq:growth_E2}
    U_{\mathrm{E}}\sim \exp\left(\omega_\mathrm{p} t\right).
\end{equation}

\begin{figure}
    \centering
    \includegraphics[width=1\columnwidth, trim={22mm 5mm 22mm 0mm}, clip]{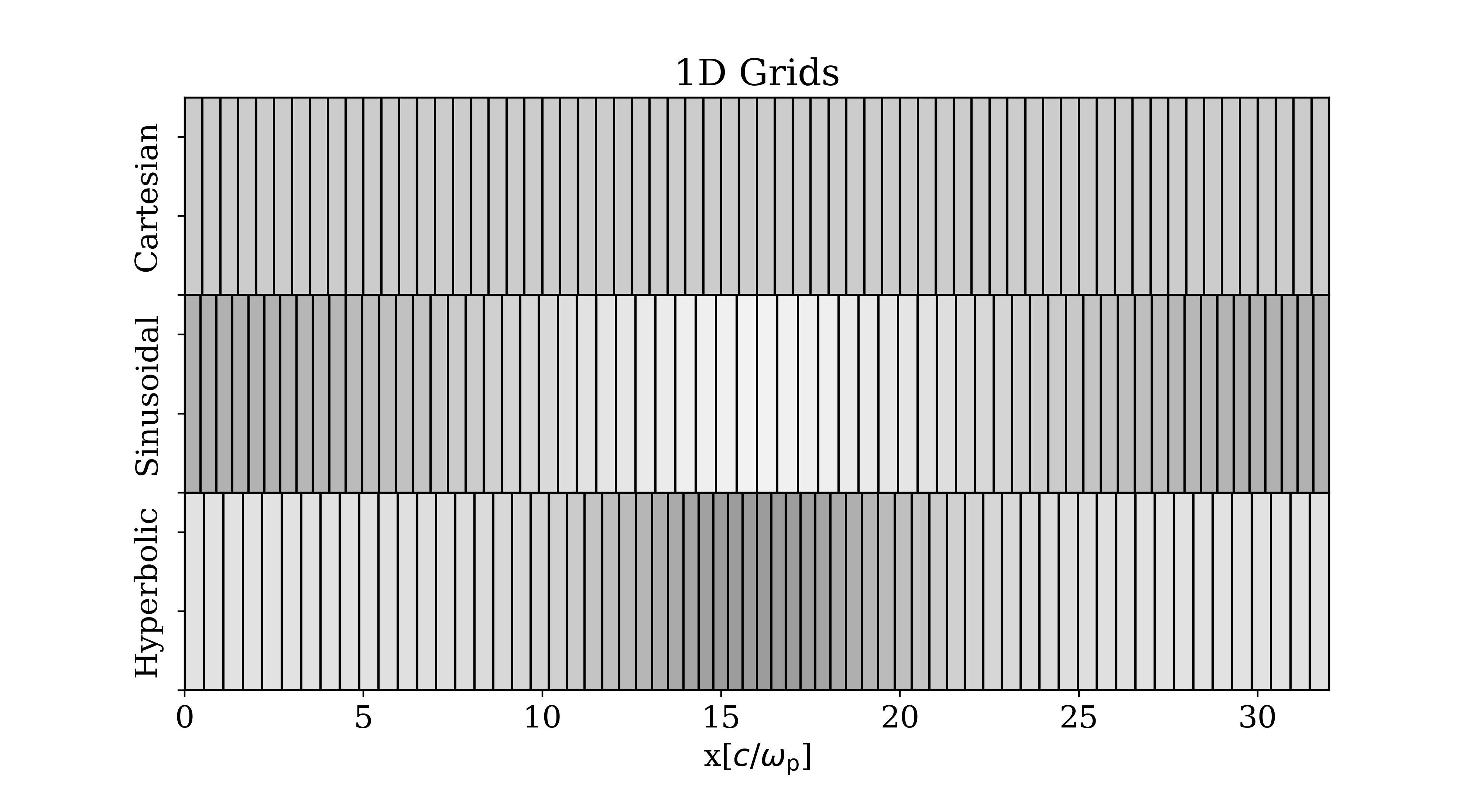}
    \caption{The three one-dimensional grids used in the study of the two-stream and Weibel instabilities: from top to bottom, Cartesian (i.e.\ the usual grid employed in standard PIC), sinusoidal, and hyperbolic-tangent mapping functions. They are shown here for a 64-cell grid with the grays-cale indicating the local grid density.}
    \label{fig:1Dgrid}
\end{figure}

To test the capability of our code to handle curvilinear coordinates, we employed two nonuniform, one-dimensional grids alongside a standard Cartesian grid used as a reference. The first curvilinear grid has a small sinusoidal perturbation of one period along the domain, with mapping function
\begin{equation}
    \xi(x) = x + \varepsilon L_x \sin\left(\frac{2\pi x}{L_x}\right),
\end{equation}
where $L_x$ the length of the domain and $\varepsilon$ parametrizes the strength of the perturbation, which we set to $\varepsilon=0.02$.
The second grid has a perturbation generated by a hyperbolic tangent, producing a grid with a $\sech^2$ profile of cell density. A high density peak is present near the center of the domain; the mapping function is
\begin{equation}
    \xi(x) = \frac{L_x- 2 \varepsilon L_x}{L_x}x - \varepsilon L_x \tanh\left[\frac{1}{w}\left(\frac{L_x}{2} - x\right)\right] + \varepsilon L_x,
\end{equation}
where $w$ determines the width of the high-density region, which we set to $4$ with $\varepsilon=0.04$. These two perturbed grids are constructed to ensure the total system length $L_x$ is maintained. This is necessary to ensure a fair comparison to the Cartesian reference grid. The parameters used in the sinusoidal and hyperbolic grids are chosen such that the ratio of the cell-width between the smallest and largest cells is $\sim 0.75$. This provides a meaningful deviation from the Cartesian case, while still allowing the same time steps without breaking the constraint $c\Delta t / \Delta x < 1$ mentioned in Section~\ref{sec:discr}. Figure \ref{fig:1Dgrid} depicts the three grids used in this test for a resolution of 64 cells. The gray-scale is an indication for local cell density, with darker colors indicating higher density. The grids used in the simulations had $N_x = 2048$ cells with a domain length of $32c/\omega_{\mathrm{p}}$, time step $\Delta t = 0.00390625\omega_{\mathrm{p}}^{-1}$ and $ppc = 144$ particles per cell.

\begin{figure*}
    \centering
    \includegraphics[width=0.9\textwidth, trim={10mm 10mm 0 0}, clip]{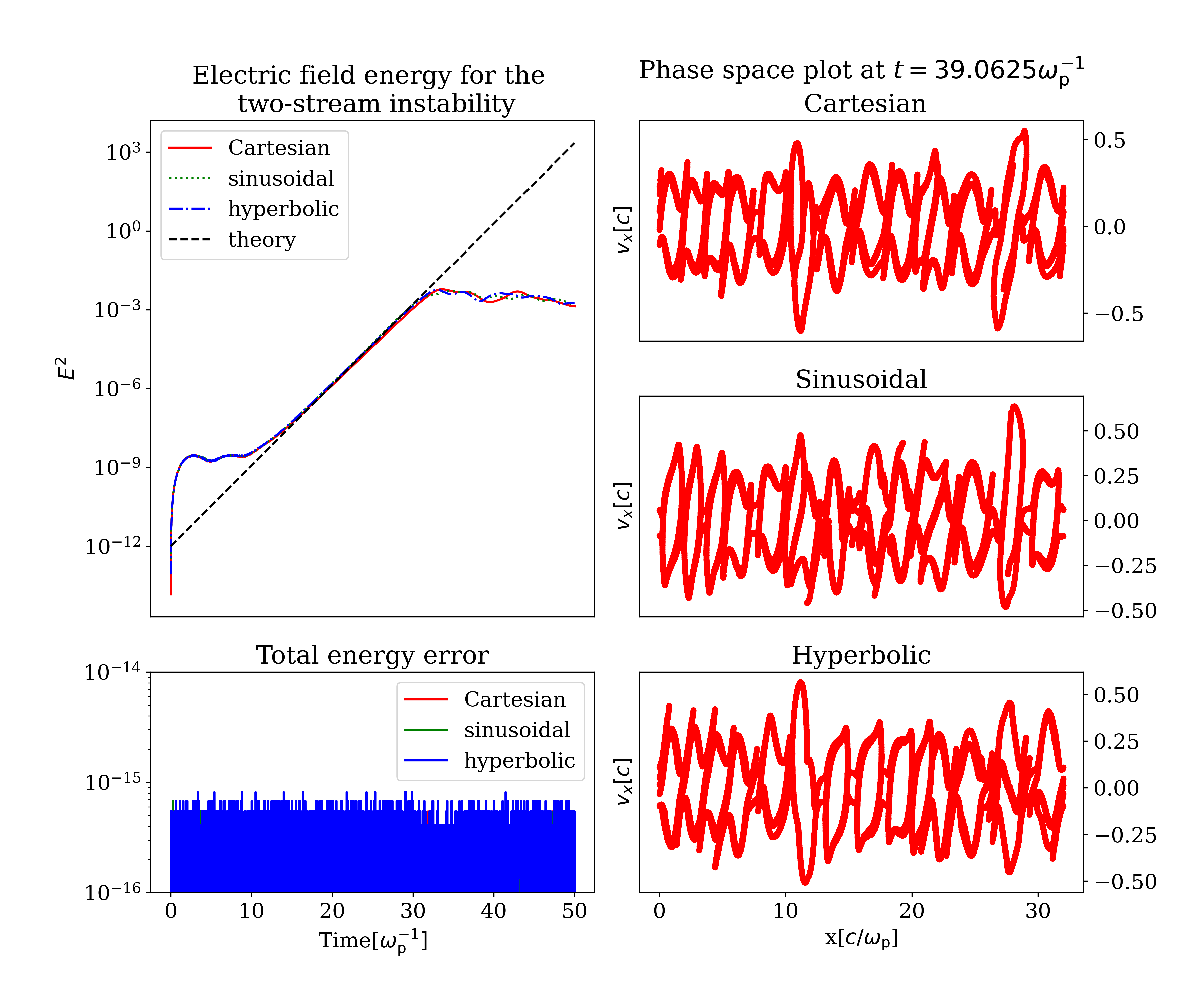}
    \caption{Top left: Evolution of the electric-field energy for the different grids as well as the theoretical growth rate associated with the linear-growth phase of the two-stream instability. Bottom left: Evolution of the relative error in the total energy for the different grids. Right: Snapshots of the $(x,v_x)$ phase space during the nonlinear stage for the different grids.}
    \label{fig:CS}
\end{figure*}

Our results for this test are shown in Figure~\ref{fig:CS}. The top left panel shows the evolution in time of the electric-field energy for the three distinct grids in one plot. It shows good agreement between the three cases as well as close adherence to the theoretical value of the growth rate outlined in eq.~\eqref{eq:growth_E2} during the linear-growth phase of the instability. The bottom left panel plots the time history of the relative error in the total energy, which confirms energy conservation to machine precision for all three cases. Lastly, the three panels on the right show a snapshot of the position-velocity phase space $(x,v_x)$, where we observe the formation of holes that are typical for the nonlinear phase of this instability. These results give a good indication that the method is capable of faithfully reproducing electrostatic effects in a nonuniform, one-dimensional setup while maintaining exact energy conservation. 

\subsection{Weibel Instability}
To test our method in a more general electromagnetic case, we study the Weibel (i.e.\ filamentation) instability. The setup is almost equivalent to the two-stream instability: $N_x = 2048$, $L_x = 32c/\omega_\mathrm{p}$, $\Delta t = 0.00390625\omega_\mathrm{p}^{-1}$ and $ppc = 144$ uniformly distributed and divided between the two electron populations.
The difference here is the direction of the beam drift velocity, which is now perpendicular to the domain's direction i.e. $v_y = v_d = \pm 0.2c$, and similarly, the thermal velocity only has a component along the perpendicular direction as well, with magnitude $v_\mathrm{th} = 0.001c$. The Weibel instability is excited by current separation generated by the Lorentz force in a perturbed magnetic field when acting on the two counterstreaming beams. The resulting current filaments will reinforce the initial magnetic-field perturbations. This leads to an exchange of energy between particles and magnetic fields. The magnetic-field energy in fastest-growing unstable mode will evolve as
\begin{equation}
    U_{\mathrm{B}} \sim \exp(2 \, v_{\mathrm{d}} \, t)
    \label{eq:weibel_gr}
\end{equation}
during the linear phase of the instability. 

\begin{figure*}
    \centering
    \includegraphics[width=0.9\textwidth, trim={10mm 10mm 0 0}, clip]{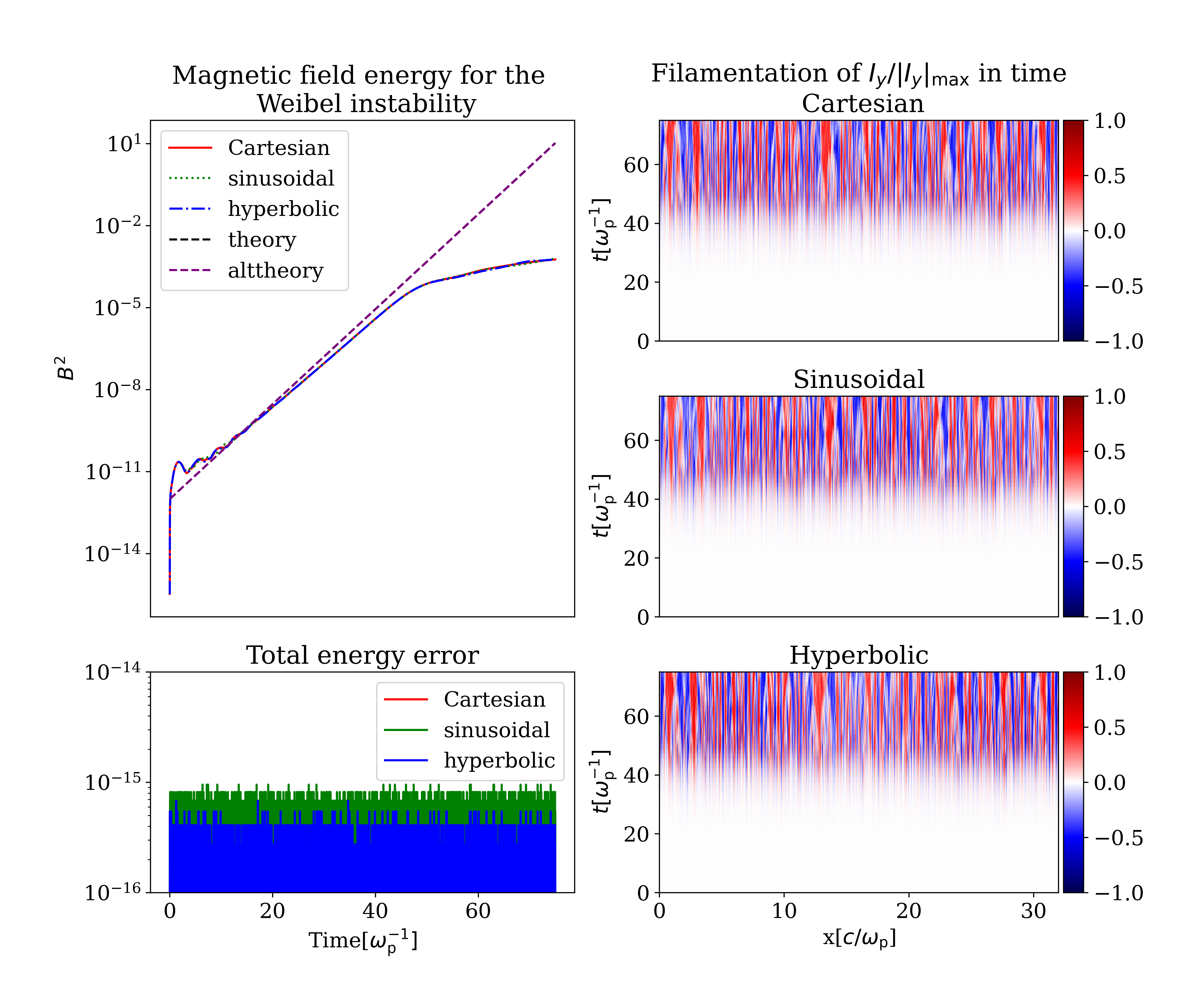}
    \caption{Top left: Evolution of the magnetic-field energy for the different grids as well as the theoretical growth rate associated with the linear-growth phase of the Weibel instability. Bottom left: Evolution of the relative error in the total energy for the different grids. Right: Formation and growth of current filaments in the domain with evolving time for the different grids.}
    \label{fig:Weibel}
\end{figure*}

Figure~\ref{fig:Weibel} shows the results of the Weibel instability tests. The upper-left panel shows the evolution in time of the magnetic-field energy for all grids. For all cases, we observe a slight discrepancy between the simulated and theoretical value of the growth rate during the linear instability phase. However, the results of the different grids are in strong agreement with each other.
The slight discrepancy with the theoretical growth rate could be attributed to different factors: for example,
the finite spatial resolution prohibits resolving large wave numbers which would otherwise increase the observed growth rate, potentially agreeing better with the theoretical maximum growth rate that corresponds to $k\to\infty$; in addition,
noise induced by the motion of the limited number of particles drowns out small waves that might otherwise contribute to the instability.
At any rate, the discrepancy is very small and the results appear robust in terms of agreement between different cases we run.
The lower-left panel shows again the evolution of the relative energy error, indicating conservation of energy to machine precision for all three grid setups. The panels on the right-hand side of the Figure show the current filaments forming and growing throughout the domain as time increases. Considering these results, the method appears  capable of correctly reproducing fully electromagnetic effects while conserving energy to machine precision, at least in this simple one-dimensional test case. 

\subsection{GEM reconnection challenge}
The last test case presented in this paper is the Geospace Environmental Modeling (GEM) reconnection challenge. This is a well-known two- (or three-)dimensional setup to study magnetic-reconnection events \citep{GEM}. We adopt a two-dimensional $xy$ geometry where the $B_x$ field undergoes a sign reversal along the $y$-axis, creating a current sheet in the inversion region. For simplicity, we adopt periodic boundary conditions, initializing two successive inversions to respect periodicity. The initial setup of the magnetic field is
\begin{equation}
    B_x = B_0 \left[1-\tanh\left(\frac{y-L_y/4}{\delta}\right) +\tanh\left(\frac{y-3 L_y/4}{\delta}\right)\right].
\end{equation}
To respect the $c\grad\times\bb{B}=4\pi \bb{I}$ condition, we initialize a drifting electron and positron population in the current sheets with a number density of 
\begin{equation}
    n = n_0 \left[\sech^2\left(\frac{y-L_y/4}{\delta}\right) +\sech^2\left(\frac{y-3 L_y/4}{\delta}\right) \right].
\end{equation}
Here $B_0$ and $n_0$ are the peak magnetic field strength and density in the current sheet respectively, and $\delta$ is the width of the current sheet, which we choose equal to $0.5 c/\omega_\mathrm{p}$.

At initialization, the thermal pressure of the particles in the current sheet must be able to withstand the magnetic pressure to ensure the structure neither collapses nor expands, i.e.\ for pressure equilibrium
\begin{equation}
    \frac{B^2_0}{8\pi} = 2 n k_{\mathrm{b}} T,                     
\end{equation}
with $k_{\mathrm{b}}$ the Boltzmann constant and $T$ the plasma temperature. The factor $2$ is introduced to include both electron and positron contribution to the density. By equating $k_{\mathrm{b}} T$ with $m v^2_{\mathrm{th}}$, a constraint for the thermal speed in the current sheet can be found,
\begin{equation}
    \frac{v_{\mathrm{th}}}{c} = \frac{B}{c\sqrt{16\pi n m}}.
\end{equation}
Note that we do not manually perturb the initial equilibrium, so that tearing modes eventually causing reconnection are only excited by random particle noise.

As in the 1D cases, a Cartesian grid is used to establish a baseline to which we compare nonuniform-grid results. The Cartesian grid has dimensions $L_x = 16c/\omega_{\mathrm{p}}$ by $L_y=32c/\omega_\mathrm{p}$ with $N_x = 160$ by $N_y = 320$ cells and $ppc = 16$ particles per cell. The time step was set to $\Delta t = 0.025\omega_\mathrm{p}^{-1}$ for $nt = 30000$ steps, for a total time of $750\omega_\mathrm{p}^{-1}$. This satisfies the convergence constraint $c\Delta t/ \Delta x < 1$.
 
To test the behavior of our code with curvilinear coordinates, we construct a nonuniform grid such that we achieve the same density of computational cells (i.e.\ the same numerical resolution) near the current sheets, while significantly reducing the grid resolution in the upstream regions. This naturally results in fewer total grid cells than in the uniform Cartesian case. This grid setup enables us to resolve the reconnection region with the same precision of the Cartesian setup while greatly reducing the total number of cells in regions of no physical interest. The mapping function for this grid was created using piecewise linear functions that were connected using hyperbolic tangents as an approximation for the Heaviside step function,
\begin{equation}
    \xi(x) = x,
\end{equation}
\begin{equation}
    \eta(y) = \frac{L_y}{p_\mathrm{max}} \left(y - \sum_{n=1}^4g_n(y)\right),
\end{equation}
where $g_n(y) = ((r_n y + p_n) - (r_{n-1} y + p_{n-1})) \tanh(s  (y - b_n))$. Here, $r_n$ is the cell-density ratio with respect to the low-density upstream regions. This is set to $1$ in the upstream regions and to a desired larger ratio for the current-sheet regions. $s$ determines the sharpness of the transition, with larger values more closely resembling the Heaviside function and lower values creating a smoother transition region. This value was set to $s=5$ for all grids. The $b_n$ terms determine the break points i.e. the $y$-coordinate where the piecewise function switches from one linear function to the next. These breakpoints therefore determine the points of transition between high and low density regions. These are set at $L_y/8$, $3L_y/8$, $5L_y/8$ and $7L_y/8$ respectively. The $p_n$ factors are the offsets of the individual linear functions, which are fully constrained by the previous parameters. Finally, $p_\mathrm{max}$ is the maximum value reached by the piecewise function, and is introduced such that $\eta\in[0,L_y]$. 

We employed two nonuniform grids, the first with a density ratio $r=5$ and number of cells in the $y$-direction $N_y = 192$, and the second with ratio $r=10$ and $N_y = 176$. The number of cells in the $x$-direction remains unchanged  from the Cartesian case ($N_x = 160$). By design, the curvilinear grid has the same physical size $L_x=16c/\omega_{\mathrm{p}}$ and $L_y=32c/\omega_{\mathrm{p}}$ of the Cartesian grid. Furthermore, the time step and number of iterations were deliberately set to be identical in both cases to again satisfy the convergence constrained, since the smallest $\Delta x$ is also unchanged. 

\begin{figure*}
    \centering
    \includegraphics[width=1\textwidth, trim={35mm 10mm 25mm 0}, clip]{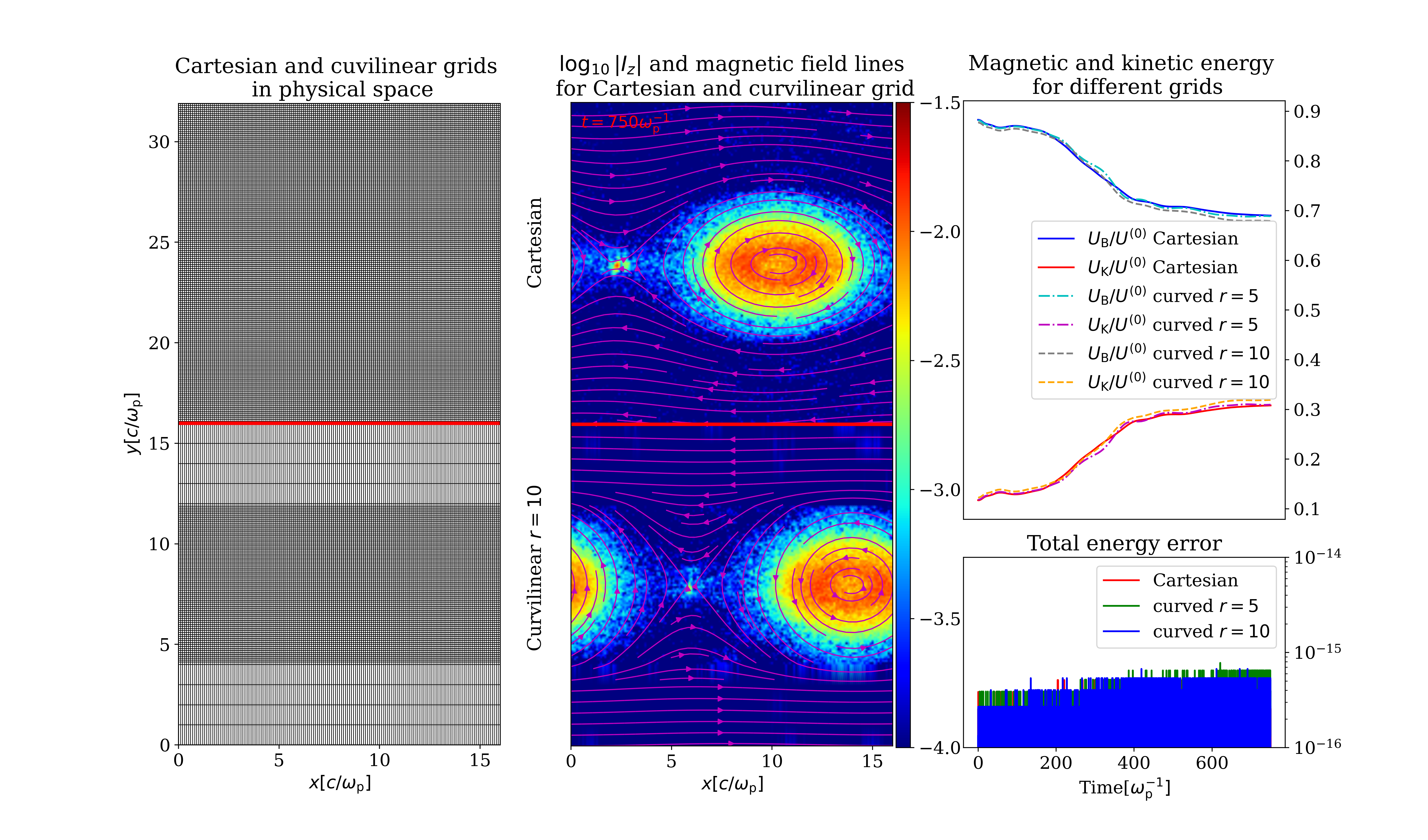}
    \caption{Left: Comparison of the uniform Cartesian grid (top) and curvilinear grid with $r=10$ (bottom) which has a reduced resolution in the upstream regions, while maintaining the same resolution in the current sheets that was used in the Cartesian grid. The magnetic islands formed in the top and bottom halves are not collocated since their positions are randomized by the starting values of the particle properties. Middle: Comparison of a snapshot of the out-of-plane current and the in-plane magnetic-field lines for the aforementioned grids. Top right: Evolution of the magnetic-field energy and kinetic energy for the different grids. Bottom right: Evolution of the relative error in the total energy for the different grids.}
    \label{fig:GEM}
\end{figure*}

Our results are shown in Figure~\ref{fig:GEM}. The left panel shows the Cartesian grid in the top half of the domain and the curvilinear grid with density ratio $r=10$ in the bottom half. In the middle panel we show a comparison of the out-of-plane current $\log_{10}|J_z|$ and the in-plane magnetic-field lines between the Cartesian grid (top) and the curvilinear grid with density ratio $r=10$ (bottom) at time $t = 750\omega_{\mathrm{p}}^{-1}$. Although it is clear that the upstream regions have considerably lower resolution in the curvilinear case, the reconnection regions of interest look very similar. Since the exact position of the formed X-points and magnetic islands is due to the random initialization of the particle properties, they appear in different places in the top and bottom current sheet in the domain. In the top right panel, we plot total magnetic and kinetic energies for all cases, to show the energy exchange between magnetic fields and particles over time. The initial magnetic field contains the majority of the energy, and as the magnetic fields reconfigure into a more relaxed state via reconnection, the particles gain energy and get accelerated. The bottom right panel shows the evolution in time of the relative error in the total energy, which is again conserved to machine precision, supporting our findings from the 1D tests and the theory in Appendix~\ref{apx:nrg}. 

\section{Conclusions}\label{sec:discussion}
In this paper, a method is presented to adapt a fully implicit PIC method to work with arbitrary coordinate systems, including curvilinear coordinates, while maintaining exact (i.e.\ to machine precision) energy conservation. While previous methods have been published for reduced equations, like the Darwin approximation (\citealt{chacon_2016}), this is the first time to our knowledge when this is achieved for a PIC with the full Maxwell equations. This is achieved by changing the coordinates to a system where the grid becomes logically rectangular. To change coordinates we introduce the metric tensor, Jacobian matrix, and contravariant curl operator into Maxwell's equations for the PIC algorithm. The particle equations of motion are solved in Cartesian coordinates, to avoid the complexity of calculating the metric and Jacobian at each particle position and at each iteration.

To validate the method, we have presented three different test cases. The two-stream instability and Weibel instability were tested in one spatial dimension and the GEM reconnection challenge was tested in two dimensions. The  tests runs with nonuniform (curvilinear) grids were compared to the results obtained with a uniform Cartesian grid, taken as a baseline. Two nonuniform grids were tested in the one-dimensional cases, both generated by perturbing the baseline Cartesian grid: the first with a sinusoidal perturbation and the second with a hyperbolic-tangent perturbation. The nonuniform grid in the GEM challenge was designed to greatly reduce the total number of grid cells, while maintaining the same spatial resolution of the uniform grid around the current sheets, to properly resolve reconnection dynamics while underresolving the unimportant upstream regions. Comparisons were performed on global quantities such as individual energy components and the total energy of the system. 

We demonstrated that our new method achieves energy conservation to machine precision in all three test cases with both Cartesian and curvilinear grids. The two-stream instability modeled with curvilinear grids shows a growth rate of the electric-field energy which closely matches both the (Cartesian uniform) baseline and the theoretical growth rate. The Weibel instability also shows excellent alignment of the magnetic-field energy to the Cartesian baseline.
In our nonuniform-grid runs, we observed that  the 2D GEM challenge produced reconnection events inside the current-sheet region as expected of this setup, despite the nonuniform grid having only $55-60\%$ the total number of cells with respect to the uniform Cartesian run. Reconnection can be observed in the in-plane magnetic-field line rearrangement, which shows the formation of magnetic islands, and in the exchange between magnetic and particle energy, showing a conversion of magnetic-field energy into kinetic energy as particles get accelerated during the reconnection process. 

Although our method possesses desirable properties (exact energy conservation and grid adaptability to complex geometries), ample ground is left for further improvements, which we will pursue in future work. Unlike the energy, the total charge is not conserved, since $\grad \cdot \bb{E} = 4\pi\rho$ is not enforced in our implementation. However, as shown in other works (e.g.\ \citealt{charge_conserving_pic_2019}), a charge-conserving adaptation of fully implicit PIC methods is possible \citep{chen2011energy}, and could be implemented in our code in the future. Another shortcoming is represented by the convergence of the nonlinear iteration, on which our method is based, which is not guaranteed; from our experiments, we observed that a sufficiently high spatiotemporal resolution must be enforced to avoid nonconvergence issues, which could break the temporal iteration. In many practical cases however, this will not pose a problem since a fine spatiotemporal resolution is required anyway, to resolve the physics of interest. In our tests, convergence (i.e.\ an absolute error below a tolerance $\sim10^{-14}$) was always attained with $\sim10$ nonlinear iterations; hence, the computational cost in our example simulations was approximately a factor 10 larger than a standard explicit PIC method, since each nonlinear iteration roughly corresponds to one explicit time cycle. The greater computational cost of our method is however counterbalanced by numerous advantages, as discussed below. In addition, speeding up convergence is in principle possible with more advanced strategies: for example, \cite{Chen_2014} have shown that by preconditioning the nonlinear solver the average number of iterations can be drastically reduced, and the spatiotemporal resolution constraints can be relaxed.

By combining a curvilinear grid with an exactly energy-conserving PIC method, we have achieved a unique set of features with numerous potential applications. In general, nonuniform grids are especially well suited for setups with a small region of interest within a much larger domain, or setups where the bulk of the plasma is contained within a spatial region with a nontrivial shape. Meanwhile, energy conservation is important in many problems where energy conversion from one type to another occurs and has significant impact on the plasma dynamics, e.g.\ reconnection events or instabilities (see \citealt{Markidis_2011} and references therein). In several cases, the combination of energy conservation and custom grids can be of high interest: in fusion plasma physics, our method could be used in simulations of the scrape-off layer (i.e.\ the plasma layer close to the separatrix) of a Tokamak plasma (e.g.\ \citealt{Ohtsuka_1978,SOL_review,SOL,Fundamenski_2007}). The nonuniform grid can be tailored to match the geometry of the boundary layer, greatly reducing the total computational cost. Another example is the expanding solar wind (e.g.\ \citealt{Innocenti_2019,Innocenti_2020,Bott_2021,Micera_2021}): here, a large domain, which expands in time as the solar wind travels, makes it difficult to employ a standard Cartesian box while maintaining sufficient resolution throughout the domain. By using a grid that mimics the shape of the expanding solar wind, with a high resolution closer to the Sun and decreasing grid density further out (where less resolution is required), the computational cost of this problem can be significantly reduced. Conversely, when considering compression-driven dynamics in plasmas, a compressing box has been used in the past in the context of simulations of the intergalactic medium (e.g.\ \citealt{Sironi_2015}), which could similarly be replaced with our nonuniform grids. 
Other solar wind-related phenomena whose modeling could benefit from a custom grid are switchbacks. Switchbacks are magnetic structures in the solar wind characterized by a typical S-shape of the plasma stream, and may be related to interchange reconnection events and particle scattering and acceleration (e.g.\ \citealt{Drake2021,Wyper_2022}). With our method, a curved grid can be fitted to this S-shaped stream, reducing the total grid resolution and cutting out the surrounding plasma that might be of little interest. This might open up the possibility to a fully kinetic study of the driving phenomena which was thus far unfeasible. 
In higher-energy plasma scenarios, our method could also find applicability e.g.\ for general-relativistic plasma simulations. Recent works have simulated plasma accretion around compact objects such as black holes (e.g.\ \citealt{parfrey_2019,Bransgrove_2021,Crinquand_2022,El_Mellah_2022,Galishnikova_2023}), where plasma flows in structures such as accretion disks and jets. These works present large-scale simulations utilizing explicit codes, usually implemented with one specific (four-dimensional) metric corresponding to the particular physical case in question. With the method presented in this work (modified by adding a time component of the metric and relativistic effects; see e.g.\ \citealt{bacchini2019fipic,bacchini2023}), we can simultaneously avoid the aforementioned downsides of explicit codes (i.e.\ numerical instability and lack of energy conservation) and generalize such simulations to any arbitrary metric tensor that might be of interest, opening the possibility for more physical cases to be studied.

We foresee several potential developments for the future. The logical next step in this line of work is to implement the method in a production-ready, parallel infrastructure such as those currently employed for implicit-PIC simulations (e.g.\ \textsc{iPic3D, ECsim}). This will open the door to larger-scale simulations outside the capabilities of our current test implementation in Python. The most obvious candidate would be the \textsc{ECsim} code (\citealt{Lapenta_Ecsim}), since it shares (by design) a very similar discretization of our governing PIC equations. This step will be carried out in the future. 
It would also be of great interest to consider novel architectures such as GPUs. The traditional bottlenecks in PIC codes are the particle gathering and deposition steps. With the increasing popularity of GPU-accelerated codes, these bottlenecks could be mitigated, since the gathering and deposition steps are well suited to parallelization (e.g.\ \citealt{Joseph_2011,DECYK2014708,Vasileska_2021}). 
The field solver, on the other hand, whether implemented on CPUs or GPUs, requires global communication of grid quantities, which hinders scaling behavior to a large number of cores, so it becomes imperative to further reduce the cost of the field-solution step through other means. With the method described in this paper, this can be achieved by optimizing the grid and thus reducing the total number of cells, which is the main driving factor of the field solver's computational cost.

\appendix

\section{Formal proof of energy conservation}
\label{apx:nrg}
Throughout this Appendix the time-level notation has been shortened for improved readability by dropping the ubiquitous $n$, thus only writing the offset from this arbitrary time level. For example $(1/2)$ is equivalent to $(n+1/2)$.

Starting from Maxwell's equations~\eqref{eq:max1}--\eqref{eq:max2}, we contract both sides with $B_{\mu}{}_g^{(1/2)}$ and $E_{\mu}{}_g^{(1/2)}$ respectively:
\begin{equation}
    \frac{c}{J_g}(B_{\mu}\epsilon^{\mu \nu \kappa}\partial_{\nu} E_{\kappa})_g^{(1/2)} = - (B_{\mu}\partial_t B^{\mu})_g^{(1/2)}, \label{eq:apx1}
\end{equation}
\begin{equation}
    \frac{c}{J_g}(E_{\mu}\epsilon^{\mu \nu \kappa}\partial_{\nu} B_{\kappa})_g^{(1/2)} = (E_{\mu}\partial_t E^{\mu})_g^{(1/2)} + 4\pi (E_{\mu} I^{\mu})_g^{(1/2)}.\label{eq:apx2}
\end{equation}
Subtracting eq.~\eqref{eq:apx1} from eq.\eqref{eq:apx2} and rearranging  terms gives
\begin{equation}
\begin{split}
    &(E_{\mu}\partial_t E^{\mu})_g^{(1/2)} + (B_{\mu}\partial_t B^{\mu})_g^{(1/2)} \\ =\,\, &\frac{c}{J_g}(E_{\mu}\epsilon^{\mu \nu \kappa}\partial_{\nu} B_{\kappa})_g^{(1/2)} - \frac{c}{J_g}(B_{\mu}\epsilon^{\mu \nu \kappa}\partial_{\nu} E_{\kappa})_g^{(1/2)} - 4\pi (E_{\mu} I^{\mu})_g^{(1/2)}.
\end{split}
\end{equation}
On the left-hand side, the time derivatives are discretized as described in Section~\ref{sec:discr} and half-integer timestep values are substituted with a temporal linear interpolation, producing
\begin{equation}
\begin{split}
    &\frac{E_{\mu}{}_g^{(1)} + E_{\mu}{}_g^{(0)}}{2}\frac{E^{\mu}{}_g^{(1)} - E^{\mu}{}_g^{(0)}}{\Delta t} + \frac{B_{\mu}{}_g^{(1)} + B_{\mu}{}_g^{(0)}}{2}\frac{B^{\mu}{}_g^{(1)} - B^{\mu}{}_g^{(1)}}{\Delta t} \\ =\,\, &\frac{c}{J_g}(E_{\mu}\epsilon^{\mu \nu \kappa}\partial_{\nu} B_{\kappa})_g^{(1/2)} - \frac{c}{J_g}(B_{\mu}\epsilon^{\mu \nu \kappa}\partial_{\nu} E_{\kappa})_g^{(1/2)} - 4\pi (E_{\mu} I^{\mu})_g^{(1/2)}.
\end{split}
\end{equation}
Summing over all grid points and multiplying with $J_g\Delta\xi\Delta\eta/4\pi$ while working out the products on the left-hand side results in 
\begin{equation}
\begin{split}\label{eq:apx_long}
    &\sum_{g}  \frac{J_{g}\Delta\xi\Delta\eta}{\Delta t} \left(  \frac{(E_{\mu}E^{\mu})_{g}^{(1)} + (E_{\mu}E^{\mu})_{g}^{(0)}}{8\pi} + \frac{(B_{\mu}B^{\mu})_{g}^{(1)} + (B_{\mu}B^{\mu})_{g}^{(0)}}{8\pi}\right) \\ = \,\,
    &\sum_{g} \frac{c\Delta\xi\Delta\eta}{4\pi} \left(E_{\mu}\epsilon^{\mu\nu\kappa}\partial_{\nu} B_{\kappa} - B_{\mu}\epsilon^{\mu\nu\kappa}\partial_{\nu} E_{\kappa}\right)_{g}^{(1/2)} 
    - \sum_{g} J_{g}\Delta\xi\Delta\eta (E_{\mu}I^{\mu})_{g}^{(1/2)}.
\end{split}
\end{equation}
On the left-hand side we can now recognize the definition of the field energies as described by eqs.~\eqref{eq:Ue}--\eqref{eq:Ub}. The first term on the right-hand side is the divergence of the Poynting-flux vector; this discrete integral (expressed by the sum over all grid points $g$) vanishes under periodic boundary conditions. This can be easily shown for the one-dimensional case by expanding the sum,
\begin{equation}
\begin{split}
    \sum_{i=0}^N &-E_{\eta  (i+1/2)} (B_{\zeta(i+1)} - B_{\zeta(i)}) + E_{\zeta  (i+1/2)} (B_{\eta(i+1)} - B_{\eta(i)})
    \\ &+ B_{\eta  (i)} (E_{\zeta(i+1/2)} - E_{\zeta(i-1/2)}) - B_{\zeta  (i)} (E_{\eta(i+1/2)} - E_{\eta(i-1/2)}) = 0,
\end{split}
\end{equation}
where every term can indeed be canceled out if the 0-th and $N$-th grid points are identical. This argument can easily be extended to two and three dimensions as well. The right-hand side in eq.~\eqref{eq:apx_long} thus reduces to the second term and can be used to denote the change in field energy between two subsequent time levels,
\begin{equation}\label{eq:part1}
    \frac{\Delta U_\mathrm{F}}{\Delta t} 
    = - \sum_{g} J_{g}\Delta\xi\Delta\eta (E_{\mu}I^{\mu})_{g}^{(1/2)},
\end{equation}
where the remaining term represents energy exchange between fields and particles.

We now have to verify that this term is identical to the energy-exchange term obtained from the particle equations. To do so we start from eq.~\eqref{eq:newt2disc} and contract both sides with $v_{i}{}_p^{(1/2)}$, 
\begin{equation}
    m_{p}v_i{}_{p}^{(1/2)}\frac{v^i{}_{p}^{(1)} - v^i{}_{p}^{(0)}}{\Delta t} = q_{p} \left(v_{i}E^{i} + v_{i}\epsilon^{ijk}v_{j}B_{k}\right)_{p}^{(1/2)}.
\end{equation}
In the last term on the right-hand side, we can recognize $v_{i}\epsilon^{ijk}v_{j}B_{k} = 0$ since this is the dot product of $\bb{v}$ with the cross product between itself and $\bb{B}$.
Like before, we rewrite the half-integer timesteps as temporal averages in the left-hand side, and we take the sum over all particles $p$ to get
\begin{equation}
    \sum_p \frac{m}{\Delta t} \frac{(v_{i}v^{i})_p^{(1)} - (v_{i}v^{i})_p^{(0)}}{2} = \sum_p q_p (v_iE^i)_{p}^{(1/2)}.
\end{equation}
We can recognize the definition of the kinetic energy as shown in eq.~\eqref{eq:Uk} and use the expression for interpolated quantities (eq.~\eqref{eq:grid_to_part}) in the right hand side:
\begin{equation}
    \frac{\Delta U_\mathrm{K}}{\Delta t} = \sum_p q_p v_i{}_{p}^{(1/2)} \sum_g w_{gp} (j_{\nu}^{i} E^{\nu})_g^{(1/2)}.
\end{equation}
Since the interpolation weight from grid to particles $w_{gp}$ is the same as the weight from particles to grid $w_{pg}$, it is possible to switch the order of the summations,
\begin{equation}
    \frac{\Delta U_\mathrm{K}}{\Delta t} = \sum_g \sum_p q_p v_i{}_{p}^{(1/2)} w_{pg} (j_{\nu}^{i} E^{\nu})_g^{(1/2)},
\end{equation}
and by adding the missing factors we can recover the right-hand side of eq.~\eqref{eq:I} for the current density,
\begin{equation}
    \frac{\Delta U_\mathrm{K}}{\Delta t} = \sum_g J_g\Delta\xi\Delta\eta (j_{\nu}^{i} E^{\nu} j_{\mu}^{i})_g^{(1/2)} \frac{({j^{-1}}_{i}^{\mu})_g}{J_g\Delta\xi\Delta\eta} \sum_p w_{pg} q_p v^i{}_{p}^{(1/2)}.
\end{equation}
Substituting and rearranging the Jacobian factors results in 
\begin{equation}
    \frac{\Delta U_\mathrm{K}}{\Delta t} =  \sum_g J_g\Delta\xi\Delta\eta (j_{\nu}^{i} j_{\mu}^{i} E^{\nu} I^{\mu})_g^{(1/2)},
\end{equation}
where we now recognize the metric $j_{\nu}^{i} j_{\mu}^{i} = g_{\nu\mu}$ and use it to lower the index of the electric field, such that
\begin{equation}\label{eq:part2}
    \frac{\Delta U_\mathrm{K}}{\Delta t} =  \sum_g J_g\Delta\xi\Delta\eta (E_{\mu} I^{\mu})_g^{(1/2)},
\end{equation}
where the energy-exchange term with the field energy is recovered.
Combining the change in kinetic energy~\eqref{eq:part2} and field energy~\eqref{eq:part1} gives the change in total energy of the system between time steps,
\begin{equation}
    \frac{\Delta U}{\Delta t} = \frac{\Delta U_\mathrm{K}}{\Delta t} + \frac{\Delta U_\mathrm{F}}{\Delta t} = \sum_g J_g\Delta\xi\Delta\eta (E_{\mu} I^{\mu})_g^{(1/2)} -  \sum_g J_g\Delta\xi\Delta\eta (E_{\mu} I^{\mu})_g^{(1/2)} = 0.
\end{equation}
Since this vanishes exactly between any two timesteps, the method conserves the total energy $U$ exactly (to machine precision) independently of the timestep, grid spacing, and any other numerical parameter.

\section*{Acknowledgements}
\label{sec:acknow}
This work was supported by the KU Leuven Bijzonder Onderzoeksfonds (BOF) under the C1 project TRACESpace.
J.C.\ acknowledges support from the European Union Horizon 2020 project DEEP-SEA (Grant agreement 955606).
L.P.\ acknowledges support from a PhD grant awarded by the Royal Observatory of Belgium.
F.B.\ acknowledges support from the FED-tWIN programme (profile Prf-2020-004, project ``ENERGY'') issued by BELSPO.
G.L.\ acknowledges support from the European Research Council (ERC) Advanced Grant ``TerraVirtualE''.
The computational resources and services used in this work were provided by the VSC (Flemish Supercomputer Center), funded by the Research Foundation -- Flanders (FWO) and the Flemish Government -- department EWI.


\bibliographystyle{aasjournal}
\end{document}